\documentclass[a4paper,11pt]{article}
\usepackage{jcappub} 
\usepackage{lineno}
\usepackage{framed}
\usepackage{color}
\usepackage{xcolor}
\usepackage{multirow}
\usepackage{lscape}
\usepackage{adjustbox,lipsum}
\usepackage{bm}

\newcommand{\Mc}{\mathcal{M}}
\newcommand{\Fc}{\mathcal{F}}

\newcommand{\Gc}{\mathcal{G}}
\newcommand{\Hc}{\mathcal{H}}
\newcommand{\pr}{^{\prime}}
\newcommand{\prt}{^{\prime\prime}}

\arxivnumber{2504.xxxx} 

\title{Cosmic inflation in non-perturbative quantum gravity}

\author[a]{Alexey S. Koshelev}
\author[a]{and Abhishek Naskar}
\affiliation[a]{School of Physical Science and Technology, ShanghaiTech University,\\ 393 Middle Huaxia Road, Pudong, Shanghai 201210, China}

\emailAdd{askoshelev@shanghaitech.edu.cn, naskara@shanghaitech.edu.cn}

\abstract{
String field theory motivated infinite-derivative models lead to non-local gravity modifications which form a promising class of quantum gravity candidates. In this paper we investigate effects of non-locality on the three-point function (the bi-spectrum) during cosmic inflation. The study is done in an Einstein frame with an infinite-derivative scalar field Lagrangian minimally coupled to the Einstein-Hilbert term. A non-local generalization of the Mukhanov-Sasaki equation is derived. Infinite-derivative operators present in this equation lead to an appearance of infinitely many new background induced states in the perturbation spectrum during inflation with complex masses on top of a usual nearly massless inflaton. On contrary to a flat background such states can be classically stable in a de Sitter space-time. This helps preserving observational constraints on the scalar power-spectrum. We proceed by studying a particular configuration assuming that the generalized Mukhanov-Sasaki equation gives rise to an inflaton and one pair of new states with complex conjugate masses as perturbative degrees of freedom. The corresponding scalar bi-spectrum is computed numerically in squeezed and equilateral limits. We use the latest observational constraints on amplitude of the bi-spectrum $f_{NL}$ from Planck~2018 dataset as a guideline for possible values of masses of new emerging states. We find that $f_{NL}$ is non-trivially sensitive to the values of complex masses and this can reduce the parameter space of gravity modifications. In particular we find that the amplitude of the squeezed limit gets easily enhanced while of the equilateral limit can stay like in a local single-field model of inflation. We end up discussing open questions relevant for this class of models of inflation.
}

\begin{document}
\maketitle
\flushbottom

\section{Introduction}
\label{sec:intro}

Stelle's modification to Einstein's General Relativity by adding curvature squared terms produces a renormalizable gravity theory \cite{Stelle:1976gc}. However this modification gives rise to a ghost as expected in any higher derivative theory according to Ostrogradski \cite{Ostrogradsky:1850fid}. One possible way out of this problem is to extend Stelle's theory with infinite derivatives operators (or form-factors) which are functions of the d'Alembertian operator with constant coefficients in the Taylor series expansion. These operator functions should be analytic at zero to preserve an IR limit. This provides a possible way to avoid ghosts in the spectrum (see \cite{Koshelev2023} for more details on this construction). Essentially one ends up with non-local gravity models \cite{Krasnikov:1987yj,Kuzmin:1989sp,Tomboulis:1997gg,Modesto:2011kw,Modesto:2014lga,Modesto:2015ozb,Biswas:2011ar,Biswas:2016egy}. In order to prevent appearance of new degrees of freedom the form-factors have to be tuned such that they can be specified up to a single entire function.
This class of models was shown to be ghost-free around Minkowski background and renormalizable by power-counting \cite{Koshelev:2017ebj} raising hopes to construct a quantum gravity theory. This makes us interested in exploring different aspects of such a construction. A vast amount of study of infinite-derivative gravity models was done in relation to inflation. In particular it was shown that Starobinsky inflation can be neatly embedded in this framework \cite{Koshelev:2016xqb} producing modified tensor-to-scalar ratio and having a potential to generate large non-Gaussianities \cite{Koshelev:2020foq} (see \cite{Koshelev2023-3} for more details). 

This approach however exhibits an interesting and unusual behavior if multiple background solutions exist. Namely, conditions to remove extra degrees of freedom in one specific background do not ensure the removal of these degrees of freedom in another one. In general these extra degrees of freedom, if present, will have complex masses squared and got named as Background Induced States (BIS) \cite{Addazi:2024qcv}.
We emphasize here that appearance of different number of states around different backgrounds is specific to any infinite derivative models and not only gravity. This can be the case as long as distinct vacua are considered. Moreover, if present, the corresponding jump in the number of degrees of freedom is infinite.
However, the problem of occurrence of degrees of freedom with complex masses is not restricted to infinite derivative models only and can happen in higher but a finite number derivative setup, say in a six-order gravity \cite{Lee:1969fy,Anselmi:2017yux}. It is important to note that if complex masses appear in a theory then it's complex conjugate will also be present as the Lagrangian should result in real observables.

Being puzzling states, BIS-s attracted significant attention and different proposals of working with BIS-s exist.
One idea is to assume or impose fine-tuned initial conditions which may result in cancellation of corresponding effects \cite{Anselmi:2017yux}. Then complex mass states are not present in on-shell asymptotic states (see a recent paper \cite{Anselmi:2025uzj} comprehensively studying this and related approaches). 
Moreover issues with causality and unitarity may arise unless certain criteria for the graviton propagator \cite{Platania:2022gtt} are satisfied.
There are attempts of solving the background-dependence of the number of states in infinite derivative gravity models by considering theories with form-factors depending on the curvature \cite{Koshelev:2022olc,Koshelev:2022bvg}. This approach, however, needs more investigation as renormalizability should be reconsidered in view of new interactions. Moreover, there are no effective methods of working with models which feature form-factors with non-constant coefficients.
As a more pragmatic and straightforward approach \cite{Tokareva:2024sct} BIS-s can be admitted to the game and one then would study related new effects.
This is our departure point in the current paper.

The main focus of this study is to explore the impact of BIS-s in non-local scalar-tensor theories.
Perhaps not surprisingly, one hardly wants and expects any influence of the new states on the power-spectrum. The main reason is that the scalar power-spectrum is strongly constrained from the latest Planck observations \cite{Planck:2018jri} with the amplitude $\Delta_{S} \sim 2.1 \times 10^{-9}$ and the spectral tilt $n_s \sim 0.96$, and this can be easily and elegantly explained by a local single-field model of inflation. This means that any corrections, if present, should be highly suppressed. On the other hand new states having a complex mass squared will have either decaying or growing classical behavior \cite{Koshelev:2020fok}. The former will produce no effect for the power-spectrum while the latter will definitely clash with the observational constraints. This in the meantime gives a constraint on the possible mass values of new states which is important for the subsequent analysis:
\begin{equation}\label{mass-parabola}
	(Im(m^2))^2 < 9 H^2 Re(m^2).
\end{equation}
This relation forms a parabola-shaped border separating classically stable and unstable masses. Modes inside this parabolic region have both solutions decaying for large times while those outside have one solution growing.
Notice that the presence of the Hubble parameter in this relation implies that the classical stability of complex masses is possible on a de Sitter background but not on the Minkowski space-time.
Values of masses are determined by the shape of infinite-derivative operators. Unfortunately, it is a significantly difficult problem to find these masses on the complex plane even if an operator is given explicitly. We thus revert the question and will study constraints on possible masses of new states focusing in this paper on scalar non-Gaussianities in such models. We assume no new states with a real mass parameter as it will be a physical ghost following the Ostrogradski statement.

In general non-Gaussian signals of scalar perturbations produced during inflation contain rich physics about the inflationary paradigm. Apart from the scalar power-spectrum (two-point correlation function), the scalar bi-spectrum (three-point correlation function) is currently a very important probe of primordial physics. Observational constraints on the bi-spectrum remain weak. The bi-spectrum is characterized by its amplitude $f_{NL}$, and momentum dependence which is known as the shape function. The Planck 2018 constrains the amplitude of bi-spectrum in different shapes \cite{Planck:2019kim} as, $f_{NL}^{local} = 6.7 \pm 5.6$ in the local limit\footnote{Apart from the equilateral and orthogonal limit Planck observations constrain the bi-spectrum at the local limit which peaks at the squeezed configuration.}, $f_{NL}^{eq} = 6 \pm 66$ in the equilateral limit. These observational bounds will be significantly improved with the next generation CMB missions like CMB-S4 \cite{CMB-S4:2022ght} and CORE \cite{CORE:2016ymi}. There are also proposals to use next generation galaxy surveys and 21-cm surveys to improve the bounds on bi-spectrum \cite{Jolicoeur:2023tcu,Andrews:2022nvv,Ballardini:2019wxj,Shirasaki:2020vkk,Yamauchi:2015mja}. 
The analysis of bi-spectrum puts a strong theoretical constraint on the single-field slow-roll inflation, known as the Maldacena consistency relation \cite{Maldacena:2002vr}. It tells us that in a single-field slow-roll setup it is impossible to produce a detectable level of bi-spectrum in a squeezed limit configuration where one of the momentum of a single excitation is very small compared to other two momenta. But this theoretical constraint can be bypassed if one considers any of the following setups: quasi-single field setup \cite{Chen:2009zp,Noumi:2012vr}, multi-field inflation \cite{Garcia-Saenz:2019njm,Rigopoulos:2005xx,Pinol:2020kvw,Senatore:2010wk,Wang:2022eop,Aoki:2024uyi,Byrnes:2010ft,Kaiser:2012ak}, single-field inflation with excited initial state \cite{Holman:2007na,Kundu:2011sg,Ganc:2011dy,Ghosh:2022cny}, curvaton scenario \cite{Lyth:2002my,Kawasaki:2011pd}, non-local gravity setup \cite{Koshelev:2020foq,Koshelev:2022bvg}. Moreover, it is very much possible that other mechanisms exist.

In the present paper we work within an inflationary model in the Einstein frame consisting of the Einstein-Hilbert term for gravity and a scalar field for the inflaton which has a kinetic term with an infinite derivative form-factor and a potential term. This model can exhibit new excitations in the spectrum of perturbations with complex masses. Indeed, one can fix the infinite-derivative form-factor to generate no new states around the perturbative Minkowski vacuum. This in return will result in BIS-s around the inflationary stage.
This particular setup from a field-theoretical point of view was considered previously in \cite{Koshelev:2020fok} and was shown to be unitary at one-loop level \cite{Koshelev:2021orf} (see also a recent paper investigating amplitudes in a similar setting \cite{Buoninfante:2024ibt}).\footnote{In general models with an infinite-derivative Lagrangian for different fields appear naturally in String Field Theory \cite{Arefeva:2001ps} and a unitarity of such models was proven in \cite{Pius:2016jsl} (see also \cite{Briscese:2021mob}).}

Cosmological aspects including a computation of possible non-Gaussianities were explored in \cite{Barnaby:2007yb} assuming a scalar field Lagrangian from $p$-adic string theory.  The latter means that both the form of the non-local operator in the kinetic term and the potential are fixed. In that paper only one state which would reduce to a standard local inflaton was considered keeping the higher-order derivatives but effectively disregarding other states in the spectrum of perturbations. In our present consideration we exactly address the question of new and presumably complex mass states. Also we keep the inflaton potential as general as possible while assuming that there is a sufficiently long lasting slow-roll stage of the inflation. That is, a standard slow-roll inflation is realized.

The paper is organized as follows. In Section~\ref{sec:compute} starting from a non-local scalar field action we first compute the non-local generalization of Mukhanov-Sasaki equation, and motivate the presence of complex mass states in the spectrum of perturbations. In Section~\ref{sec:2pt}  we discuss the quantization schemes for these excitations, compute the scalar power-spectrum and discuss the effect of the complex masses. In Section~\ref{sec:3pt} we introduce non-local Wightman functions needed for the computation of the bi-spectrum in our model. In Section~\ref{sec:numerical} we numerically compute the bi-spectrum with In-In formalism and discuss constraints on the complex masses coming from the observational bounds on $f_{NL}$ with several benchmark values. The paper is concluded by Discussion~Section which summarizes the results.

\section{Inflation with an infinite derivative inflaton}\label{sec:compute}

\subsection{Non-local scalar-tensor inflation}

We start with a non-local scalar field Lagrangian minimally coupled to Einstein-Hilbert gravity,
\begin{equation} \label{eq:Lag-full}
    S = \int d^4 x \sqrt{-g} \left[\frac{M_P^2}{2}R + \frac12\varphi  (\Box-m^2) \mathcal{F}({\Box_s})^2 \varphi - V(\varphi) \right],
\end{equation}
where
\footnote{The following notations are adapted: throughout the paper we work with the spatially flat Friedmann metric $ds^2=-dt^2+a(t)^2d\vec x^2$ in four dimensions where the signature is self-evident, a `dot' denotes a derivative with respect to the cosmic time $t$, a `prime' denotes a derivative with respect to the conformal time $\tau$ defined as $dt=a(\tau)d\tau$ (we also will denote it sometimes as $\partial_0$), the covariant d'Alembertian operator is $\Box=\nabla_\mu\nabla^\mu$, where $\nabla_\mu$ is a covariant derivative adding a minus connection term for a lower index and a plus connection term for an upper index. Greek indices run $0,1,2,3$. Other notations will be specified as long as they arise.}
$M_P$ is the Planck mass, $R$ is the Ricci scalar defined through the Riemann tensor as $R=g^{\alpha\beta}R^\mu_{\alpha\mu\beta}$, $\varphi$ is a scalar field of mass $m$ (in a trivial Minkowski vacuum) which we identify as the inflaton field, $\mathcal{F}\left({\Box_s}\right)$ is a form-factor, $\Box_s=\Box/\mathcal{M}^2$ where $\mathcal{M}$ is the non-locality scale whose value will be discussed below. $V(\varphi)$ is the interaction potential for the inflaton field, i.e. $V(\varphi)$ contains only interaction terms with the powers of field grater than $2$.

The form-factor is a function analytic at zero to have a well-defined IR limit, that is $\Fc(\Box_s) =\sum\limits_{n\geq0}f_n\Box_s^n$ and $f_0=1$ to preserve a canonical normalization of the scalar field. Then we see that $\Mc\to\infty$ restores a local theory. Since we do not expect singularities for some values of momenta in a Lagrangian, we naturally expect our form-factor to be even an entire function. 
This way we can guarantee that a Minkowski vacuum corresponding to $\varphi=0$ will have no extra excitations as long as $\Fc(\Box_s)=e^{2\sigma(\Box_s)}$ where $\sigma(\Box_s)$ is an entire function. Indeed, given we count the number of degrees of freedom as the number of poles in the propagator, there will be no new poles from an exponent of an entire function because such an exponent is yet another special entire function with no zeros and thus no poles if it is inverted.
This however implies that the form-factor is obligatory an infinite-derivative, or non-local, operator.

The above invertible choice of $\Fc(\Box_s)$ allows us to canonically normalize the field $\varphi$ as follows
\begin{equation}
    \chi = \mathcal{F}(\Box_s) \varphi, ~ \varphi = \mathcal{F}(\Box_s)^{-1} \chi\equiv\Gc(\Box_s)\chi\equiv\tilde \chi,
\end{equation}
and we can rewrite action \eqref{eq:Lag-full} as
\begin{equation}\label{eq:canLag}
    S = \int d^4 x \sqrt{-g} \left[\frac{M_P^2}{2}R + \frac{1}{2} \chi \left(\Box-m^2\right) \chi - V(\tilde \chi)\right].
\end{equation}
Here the non-localities are transferred to the interaction potential as $V(\tilde{\chi})$ while the kinetic term has a canonical local form. The operator $\Gc(\Box_s)$ defined above is also an entire function by construction and we can write its Taylor series expansion as $\Gc(\Box_s) = \sum\limits_{n\geq0} g_n \Box_s^n$. Note that $\Fc(0)=1$ implies that $\Gc(0)=1$ as well.

The energy-momentum tensor of the scalar field $\chi$ becomes,
\begin{eqnarray}\label{eq:EMT} \nonumber
    T_{\mu \nu} &=& \partial_{\mu} \chi \partial_{\nu} \chi -  g_{\mu \nu}\left\lbrace \frac{1}{2} g^{\alpha \beta} \partial_{\alpha} \chi \partial_{\beta} \chi  + \frac12 m^2 \chi^2 + V(\tilde\chi)\right\rbrace  \\ \nonumber
    &+& \frac{1}{2} \sum\limits_{n=1} g_n \sum_{l=0}^{n-1} \left\lbrace \frac{1}{\Mc^2}\partial_{\mu} \Box_s^l V^{(1)}(\tilde\chi) \partial_{\nu} \Box_s^{n-1-l} \chi + (\mu \leftrightarrow \nu)  \right. \\  &-&\left.  g_{\mu \nu} \left(\frac{1}{\Mc^2}g^{\alpha \beta} \partial_{\alpha} \Box^l V^{(1)}(\tilde\chi) \partial_{\beta} \Box_s^{n-1-l} \chi + \Box_s^l V^{(1)}(\tilde\chi) \Box_s^{n-l} \chi\right)\right\rbrace.
\end{eqnarray}
Here $V^{(1)}(\lambda)$ is the derivative of the interaction potential with respect to its argument (and not with respect to a scalar field). The first line of $T_{\mu \nu}$ is similar to a local field theory but with a potential having as an argument a non-local form-factor acting on a field. The second and third lines consist of contributions unique to a non-local theory. The local case is reproduced when $\Gc(\Box_s) = 1$.
We define the first and second slow-roll parameters similar to a local case scenario as,
\begin{eqnarray}\label{eq:2ndSlow}
	\epsilon = -\frac{\dot{H}}{H^2},\quad
	\eta = \frac{1}{8 \pi G}\frac{\tilde{V}_{\chi \chi}}{\tilde{V}}\Big{\vert}_{\Gc(\Box_s)=1},
\end{eqnarray}
where $H=\dot a/a$ is the Hubble parameter, $\tilde V=\frac12m^2\chi^2+V(\tilde\chi)$ and $\tilde V_\chi$ here is a derivative with respect to a field. $G$ is Newton's constant related to Planck mass as $1/M_P^2 = {8 \pi G}$. In a local inflationary setup the slow-roll conditions imply $\epsilon,~ \eta \ll 1$ during inflation. In formula \eqref{eq:EMT} there are non-local corrections proportional to $1/\Mc^{2n}$. We consider the scale of inflation to be much lower than the non-locality scale as argued in \cite{Koshelev:2016xqb}, or in other words $H \ll \Mc$. With these considerations we can neglect terms suppressed by the non-locality scale at least at the background level and assume $\epsilon \simeq  4 \pi G \frac{\dot{\chi}^2}{H^2}$ and $\eta \simeq \frac{1}{8\pi G} \frac{\tilde{V}_{\chi\chi}}{3H^2}$. This is logic because otherwise higher-derivative effects would interfere with the inflation too much while we know that inflation can efficiently be realized as a simple single scalar field local model. This implies that a standard slow-roll solution will be a solution in our case up to corrections suppressed by inverse powers of $\Mc$.
For completeness we write down the equation of motion for field $\chi$ which takes the form
\begin{equation}\label{eq:EOM}
    \Box \chi - m^2 \chi - \Gc(\Box_s)V^{(1)} (\tilde\chi)= 0.
\end{equation}

\subsection{Scalar perturbations}
In this Section we want to compute scalar perturbations in our model (\ref{eq:Lag-full}). We consider the case of a spatially flat Friedmann metric
\begin{equation}\label{eq:metric}
    ds^2=-dt^2+a(t)^2 \delta_{ij} dx^i dx^j = a(\tau)^2 \left( -d\tau^2 + \delta_{ij} dx^i dx^j \right),
\end{equation}
where $a(t)$ (likewise $a(\tau)$) is the scale factor. Scalar perturbations for the metric and and the field $\chi$ can be written in the following way,
\begin{eqnarray}\label{eq:pertL}
	ds^2 &=& a(\tau)^2 \left\lbrace-(1 + 2 \phi) d\tau^2 - 2\partial_i \beta d\tau dx^i + \left[(1 - 2\psi)\delta_{ij} + 2 \partial_i \partial_j \gamma \right] dx^i dx^j \right\rbrace, \\
	\chi &=& \bar{\chi} + \delta \chi.
\end{eqnarray}
We implement the longitudinal gauge $\beta=\gamma=0$ in which the metric perturbations $\phi$ and $\psi$ become equivalent to the gauge-invariant Bardeen potentials as $\phi=\Phi$ and $\psi=\Psi$ and moreover the scalar field fluctuation $\delta \chi$ becomes equivalent to its gauge-invariant counterpart. From now on we will drop the bar on the background scalar field $\bar{\chi}$ and denote it as $\chi$ for brevity.

In computing perturbations it is obvious that since there is no non-local form-factor present in the gravity sector in  \eqref{eq:Lag-full} perturbations of the  Einstein tensor will be exactly like in a local theory. Instead all the non-local contributions will be coming from the scalar field energy-momentum tensor   \eqref{eq:EMT}. The local part of the energy-momentum tensor will obviously not introduce any new terms. So we focus on the new non-local terms. Those can be  perturbed carefully accounting variations of each and every d'Alembertian and the scalar field.

However, significant simplifications are coming upon accounting the fact that the background exhibits a slow-roll inflation. This means that terms where derivatives are acting on the background quantities are suppressed by slow-roll parameters. For example, term $\partial_{0}V^{(1)}(\tilde \chi) \partial_{\nu} \left(\delta\left(\Box_s\right) \chi\right)$ is the leading in the slow-roll approximation as only one time derivative acts on $V^{(1)}$. Higher derivatives acting on $V^{(1)}(\tilde \chi)$ are essentially sub-leading.
Moreover, $\delta (\Box) = -2 \Psi \Box + \frac{4}{a^2} \Psi\pr \partial_0$ and this will induce additional time derivatives on $\chi$. Therefore, the leading order contribution will be $V^{(2)}(\tilde \chi) \partial_{0}\left(\tilde\chi\right) \chi\pr \Psi\pr \simeq V^{(2)}(\tilde\chi) {\chi^\prime}^2 \Psi\pr$. As variations of d'Alembertians introduce additional slow-roll suppression we can ignore them.
Upon lengthy and tedious computations we can write two equations for $\Psi$ and $\delta \chi$ as follows,
\begin{eqnarray}\label{eq:psiEOM}
    \Psi^{\prime \prime} + k^2 \Psi^{\prime} + \Hc \Psi^{\prime}+\Hc^{\prime} \Psi -2 \frac{\chi^{\prime \prime}}{\chi{\prime}} (\Psi^{\prime}+\Hc \Psi) =  \mathcal{Z}, \\ \nonumber
    2 \Psi (\partial_{0}^2 + 2 \Hc \partial_{0}) \chi - 4 \Psi\pr \chi\pr + \delta \chi\left(\partial_{0}^2 + 2 \Hc \partial_{0} - \partial_i \partial^i\right)\delta \chi = \\ \label{eq:chiEOM}
    - a(\tau)^2\left\{m^2 \delta\chi + \Gc(\Box_s)(V^{(2)}(\tilde\chi)\Gc(\Box_s) \delta \chi)\right\},
\end{eqnarray}
with
\begin{equation*}
    \mathcal{Z} =  - 4\pi G  \frac{\chi\prt}{\chi\pr} \sum^\infty_{n=1}\frac{g_n}{\Mc^2}\sum\limits_{l=0}^{n-1} \left\{\partial_{0}(\Box_s^l V^{(1)}(\tilde\chi))\Box_s^{n-1-l}\delta\chi + \partial_{0}(\Box_s^{n-1-l} \chi) \Box_s^l (V^{(2)}(\tilde\chi)\Gc(\Box_s)\delta \chi)\right\}.
\end{equation*}
Here $\Hc=a H=a'/a$ is the conformal Hubble parameter. In the above equations we have indirectly used a perturbation of the $i\neq j$ (here $i,j=1,2,3$ are spatial indices) Einstein equation which assures that $\Phi=\Psi$ exactly like it is in a local scalar-tensor theory.
In a local limit $\Gc = 1$ and therefore $g_n=0$ for $n\geq 1$ leading to $\mathcal{Z}=0$.

Next we need to combine \eqref{eq:psiEOM} and \eqref{eq:chiEOM} to a single equation for one variable. We can proceed like in a local case by introducing a gauge-invariant variable $v = a \left(\delta\chi+\Psi \chi\pr/\Hc \right)$ known as Mukhanov-Sasaki variable. In doing so we arrive at first to the following lengthy expression,
\begin{eqnarray} \nonumber 
      \left(\partial_{0}^2 +2\Hc \partial_{0} - \partial_i \partial^i\right) \frac{v}{a} + \left(3\eta - 6 \epsilon\right)\Hc^2 \frac{v}{a} + \left\{\left(3\Hc^2 \eta - m^2 a^2\right)  \left( \Gc(\Box_s)^2 -1 \right)-\mathcal{Y}\right\} \frac{v}{a} = \\ \nonumber 
      \left\{2\frac{\mathcal{Y}}{\Hc\pr-\Hc^2}\left[a^2 \Gc(\Box_s)V^{(1)}(\tilde\chi)- \Hc \chi\pr\right]-\frac{\chi\pr}{\Hc^2}\partial_{0}\mathcal{Y}+\frac{2}{\Hc}a^2 \partial_0\left((\Gc(\Box_s)-1)V^{(1)}(\tilde\chi)\right)\right\}\Psi \\ \nonumber
      - a^2 \left\{ V^{(2)}(\tilde\chi) \left(\Gc(\Box_s)^2 - 1\right)\right\}\left(\frac{\chi\pr}{\Hc}\Psi\right) \\ \nonumber
      - \left(6 \epsilon \Hc^2 + \epsilon \mathcal{Y} \right) \frac{1}{\chi\pr} \sum\limits_{n=1} g_n\sum\limits_{l=0}^{n-1} \frac{1}{\Mc^2}\left\{\partial_0\left(\Box_s^l V^{(1)}(\tilde\chi)\right)\Box_s^{n-l-1}\left[\frac{v}{a}+\frac{\chi\pr}{\Hc}\Psi\right]  \right. \\  \left. + \partial_0\left(\Box_s^{n-l-1}\chi\right)\Box_s^{l}\left(V^{(2)}(\tilde\chi)\Gc(\Box_s)\left[\frac{v}{a}+\frac{\chi\pr}{\Hc}\Psi\right]\right)\right\},\label{eq:MSmaster}
\end{eqnarray}     
\begin{equation*}
    \text{where~}\mathcal{Y} = 4\pi G\sum_{n=1}^\infty g_n
    \sum\limits_{l=0}^{n-1}\frac{1}{\Mc^2}\partial_{0}\left(\Box_s^l V^{(1)}(\tilde\chi)\right) \partial_{0}\left(\Box_s^{n-1-l}\chi\right).
\end{equation*}
 In coming to equation \eqref{eq:MSmaster} we dropped various terms beyond the leading order in the  slow-roll approximation. In a local single-field case the last term on the left hand side of (\ref{eq:MSmaster}) and all terms on the right hand side of \eqref{eq:MSmaster} are zero. However, in a non-local scenario the form-factor introduces extra terms which are responsible for non-vanishing contributions containing $\Psi$. Fortunately we can see that all the terms containing $\Psi$  in \eqref{eq:MSmaster} are slow-roll suppressed and the leading order terms will only originate from the first line of this equation. Hence under our approximations it can be shortened to
\begin{equation}\label{eq:nlMS}
    \left(\partial_{0}^2+ 2\Hc \partial_{0} -\partial_i\partial^{i} \right) \frac{v}{a} + \left(3\eta - 6 \epsilon\right)\Hc^2 \frac{v}{a} + \left(3\Hc^2 \eta - m^2 a^2 \right)\left(\Gc(\Box_s)^2 - 1\right) \frac{v}{a}=0.
\end{equation}
We will name it a non-local Mukhanov-Saski equation. In the local case when $\Gc(\Box_s)=1$ this equation reduces to a familiar (local) form of Mukhanov-Sasaki equation
\begin{equation}\label{eq:localMS}
    v\prt - \partial_i \partial^i v -\frac{\nu^2-\frac{1}{4}}{\tau^2}v=0,
\end{equation}
where $\nu^2 = \frac94 + 9\epsilon - 3 \eta$. The quantity $v$ in the local inflation scenario is the canonically normalized scalar  related to curvature perturbation $\zeta$ as $v = z \zeta$ where $z = a \chi\pr/\Hc \sim a\sqrt{2 \epsilon}$. Notice that compared to a local single-field case when Mukhanov-Sasaki equation can be derived without any approximations here it is crucial to use both the slow-roll conditions and the  $H \ll \Mc$ condition specific to a non-local model.

Compared to  \eqref{eq:localMS},  equation \eqref{eq:nlMS} has $1/\Mc^{2n}$ corrections coming from $\Gc(\Box_s)$. While such corrections are neglected in our analysis on the background here in equation (\ref{eq:nlMS}) they appear in front of perturbations and thus cannot be easily dropped. The non-local operator present in  \eqref{eq:nlMS} which acts on $v/a$ can be written as, 
\begin{equation}\label{eq:nlOp}
    \mathcal{P}(\Box)= a^2\Box - (3\eta - 6 \epsilon)\Hc^2 - \left(3\Hc^2 \eta -m^2 a^2\right)\left(\Gc(\Box/\Mc^2)^2 - 1\right) .
\end{equation}

We remind \cite{Koshelev:2007fi} here that upon solving a linear equation of a form $\tilde{\mathcal{P}}(\Box)w=0$ where $\tilde{\mathcal{P}}(\Box)$  is an entire function with constant Taylor series coefficients we can use the fact that a general solution will be given by $w=\sum_i w_i$ where $(\Box-{\tilde m}_i^2)w_i=0$ for each $i$ and ${\tilde m}_i^2$ are roots of an algebraic equation $\tilde{\mathcal{P}}({\tilde m}_i^2)=0$. Hereafter Latin index $i$ is mostly used to enumerate perturbative excitations and is not a spatial index.
These roots can be read from the so called Weierstrass product decomposition for entire functions given by $\mathcal{P}(\Box)\equiv \prod\limits_i \left(\Box - m_i^2\right)e^{p(\Box_s)}$, where $p(\Box_s)$ is yet another entire function. Those $w_i$ form essentially a maybe infinite set of degrees of freedom.

In our case of equation \eqref{eq:nlMS} operator $\mathcal{P}(\Box)$ has strictly speaking non-constant Taylor series coefficients. However, under our approximations we can assume that the slow-roll parameters are nearly constants. Moreover, in the nearly de Sitter stage of expansion $\mathcal{H}^2\sim{a^2}\sim\frac1{\tau^2}$ and as such this factor can be taken out in the definition of operator $\mathcal{P}(\Box)$. This allows to use the just outlined method of solving the equation for $v$.
Since $\Gc(\Box_s)$ is an exponent of an entire function, operator  $\mathcal{P}(\Box)$ can not be an exponent of an entire function. Moreover it cannot be a finite degree polynomial.
Nevertheless since $\Gc(\Box_s)$ is an entire function, the operator $\mathcal{P}(\Box)$ is an entire function again and it can be decomposed using the Weierstrass product.
Thus the derived structure of the operator $\mathcal{P}(\Box)$ will result in an infinite number of degrees of freedom.

In order to proceed we take  $v$ as, $v=\sum_{i} v_i$ such that  $\left(\Box-m_i^2\right) \frac{v_i}{a} = 0$ is satisfied for every $i$. Individual equations for $v_i$ are local but the analog of mass parameter $m_i$ can be complex \cite{Koshelev:2020fok}. It is obvious that any complex $m_i$ implies a presence of its complex conjugate companion as all the coefficients in the Taylor expansion are assumed to be real. This in the meantime helps preserving the reality of the Lagrangian. In fact apart from a single real value we naturally expect all other $m_i$ to be complex as otherwise some new real mass degrees of freedom will correspond to physical ghosts.
 Values $m_i$ are determined by the shape of the form-factor. However, even if a form-factor is given explicitly it can be a very involved problem to find all the corresponding mass parameters. On the other hand the subsequent analysis does not depend on particular values of $m_i$ and this allows us to take another route. Namely, we are going to study effects of new degrees of freedom and use our results as constraints on possible form-factors. One of the important condition mentioned already in the Introduction is that complex values of $m_i$ to be classically stable should satisfy relation (\ref{mass-parabola}). We recall that a classical stability of complex masses is possible on a (nearly) de Sitter background but is not possible in the Minkowski space-time.
 
If there is no non-locality in the model then it is easy to see that only a single $v_i = v_0$ corresponding to $m_0$ exists which is the same as the solution to \eqref{eq:localMS} which we denote as $v_{MS}$. In the non-local case corrections of $\Mc^{2n}$ actually modify the $v_0$ solution, namely a mass get shifted by an additional term suppressed by $1/\Mc^2$. But in the approximation $\Mc \gg H$ it is clear that $v_0 \simeq v_{MS}$. Then as described above in addition to $v_0$ new excitations with presumably complex masses appear  which come in complex conjugate pairs and we will denote the $i$-th pair of such excitations as $v_i$ and $v_i^*$. These excitations satisfy the following equations
\begin{eqnarray} \label{eq:lMS}
    v_0^{\prime \prime} - \partial_i \partial^i v_0 - \frac{\nu_0^2 - \frac{1}{4}}{\tau^2} v_0 &=& 0, \\ \label{eq:lcMS}
   v_{i}^{\prime \prime} - \partial_k \partial^k v_{i} - \frac{\nu_{i}^2 - \frac{1}{4}}{\tau^2} v_{i} &=& 0, ~i\geq1,\\ \label{eq:lcMS1}
   v_i^{*\prime \prime} - \partial_k \partial^k v_{i}^* - \frac{\nu_{i}^{*2} - \frac{1}{4}}{\tau^2} v_{i}^* &=& 0, ~  i \geq 1,
\end{eqnarray}
here a `*' denotes the complex conjugation and the quantities $\nu_0$, $\nu_{i}$  can be written as follows,
\begin{eqnarray}
    \nu_0^2 = \frac{9}{4} + 9 \epsilon - 3\eta;~
    \nu_{i}^2 = \left(\frac{9}{4} - \frac{m_i^2}{H^2} \right). 
\end{eqnarray}  
Strictly speaking even though equation for $v_i$ and $v_i^*$ look like complex conjugate, solutions maybe not. It is an additional requirement on initial conditions which should be imposed as long as one wants to have real physical observables in such a model, energy for example.

Equations \eqref{eq:lMS}, \eqref{eq:lcMS}, \eqref{eq:lcMS1} are local equations and the non-locality of the original Lagrangian \eqref{eq:Lag-full} manifests itself through the presence of additional excitations at the perturbative level. This makes our analysis distinct from the multi-field inflation \cite{Iarygina:2023msy,Wang:2022eop,Rigopoulos:2005ae,Kaiser:2012ak,Garcia-Saenz:2019njm,Pinol:2020kvw,Senatore:2010wk}. Also our approach of treating non-locality is also more complete than the previous works \cite{Barnaby:2007yb,Lidsey:2007wa} where the non-locality was considered affecting the background only while not considering its effect on the perturbations.

\section{Two-point correlation functions}\label{sec:2pt}
Quantization of the  real $v_0(\tau,x)$ and complex conjugate pairs $v_{i}(\tau,x)$, $v_i^*(\tau,x)$ of excitations can be written as follows
\begin{eqnarray} \label{eq:quant1}
    \hat{v}_0 (\tau,x) &=& \int \frac{d^3 k }{(2\pi)^{3/2}} \left\{\hat{a}^0_k v_{0,k}(\tau) e^{i \vec{k}\vec{x}}+ \hat{a}_k ^{0\dagger}v_{0,k}^*(\tau) e^{-i \vec{k}\vec{x}} \right\}, \\ \label{eq:quant2}
    \hat{v}_i (\tau,x) &=& \int \frac{d^3 k }{(2\pi)^{3/2}} \left\{\hat{a}_k^i v_{i,k}(\tau) e^{i \vec{k}\vec{x}}+ \hat{b}_k^{i \dagger} u_{i,k}(\tau) e^{-i \vec{k}\vec{x}}\right\}, \\ \label{eq:quant3}
    \hat{v}_i^* (\tau,x)&=& \int \frac{d^3 k }{(2\pi)^{3/2}} \left\{\hat{b}_k^i u_{i,k}^*(\tau) e^{i \vec{k}\vec{x}}+ \hat{a}_k^{i \dagger} v_{i,k}^*(\tau) e^{-i \vec{k}\vec{x}}\right\}.
\end{eqnarray}
Here $v_{0,k}$, $v_{i,k}$, $u_{i,k}$ are the  corresponding mode functions, and repeated indices do not imply summation. Notice that operator $\hat{v}_i^* (\tau,x)$ is the complex conjugate of operator $\hat{v}_i(\tau,x)$ and once the solution for $\hat{v}_i(\tau,x)$ is determined, $\hat{v}_i^* (\tau,x)$ is also determined by the complex conjugation. The commutation relations for the creation and annihilation operators of the $0$-th and  $i$-th excitations are 
\begin{eqnarray}\label{eq:comm1}
    \left[\hat{a}_k^0, \hat{a}_{k^{\prime}}^{0\dagger}\right] = \delta^{(3)}(k - k^{\prime});~ 
    \left[\hat{a}_k^i, \hat{a}_{k^{\prime}}^{j\dagger}\right] = \delta^{(3)}(k - k^{\prime}) \delta^{ij};~
    \left[\hat{b}_k^i, \hat{b}_{k^{\prime}}^{j\dagger}\right] = \delta^{(3)}(k - k^{\prime}) \delta^{ij},
\end{eqnarray}
while all other commutation relations are zero.

This quantization prompts several comments. First, such a system was studied in earlier papers \cite{Yamamoto:1970gw} but the quantization prescription proposed in that paper has several features which seem to be inconsistent. For example, creation of the backward going wave in the vacuum for a complex conjugate partner. Second, in a recent paper \cite{Tokareva:2024sct} a different quantization scheme for such complex conjugate fields was proposed which seems to be consistent but its formulation resorting to purely real degrees of freedom looks more complicate then ours. In our formulation we take care of the fact that in the limit when $m=m^*$, i.e. when the mass is real, we restore a quantization procedure for a standard $U(1)$ symmetric complex scalar field as described in details in \cite{Bogolyubov:1959bfo}.

The solutions to the mode functions $v_{0,k}(\tau)$, $v_{i,k} (\tau)$ and $u_{i,k}(\tau)$ according to  \eqref{eq:lMS}, \eqref{eq:lcMS} and \eqref{eq:lcMS1} can be written as
\begin{eqnarray}
     v_{0,k}(\tau) &=& \sqrt{-\tau} \left[\alpha_0 H^{(1)}_{\nu_0}(-k \tau) + \beta_0 H^{(2)}_{\nu_0}(-k \tau)\right], \\
    v_{i,k} (\tau) &=& \sqrt{-\tau} \left[\alpha_i H^{(1)}_{\nu_i}(-k \tau) + \beta_i H^{(2)}_{\nu_i}(-k \tau)\right],\\
    u_{i,k} (\tau) &=& \sqrt{-\tau} \left[\tilde{\alpha}_i H^{(1)}_{\nu_i}(-k \tau) + \tilde{\beta}_i H^{(2)}_{\nu_i}(-k \tau)\right].
\end{eqnarray}
Here, $H^{(1,2)}_{\nu}(-k \tau)$ are Hankel functions of first and second kind respectively and $\alpha$, $\beta$, $\alpha_i$, $\beta_i$, $\tilde{\alpha}_i$ and $\tilde{\beta}_i$ are constants which can be determined by choosing the initial states of corresponding excitations. 

As standard for the inflation we assume all of the excitations to start in Bunch-Davies initial state and this yields,
\begin{eqnarray}
    \alpha_0 &=& \alpha_i = e^{i\frac{\pi}{2}\left(\nu + \frac{1}{2}\right)}\sqrt{\frac{\pi}{4}};~ \tilde{\beta}_i = e^{-i\frac{\pi}{2}\left(\nu + \frac{1}{2}\right)}\sqrt{\frac{\pi}{4}};~ \beta_0=\beta_i = \tilde{\alpha}_i = 0.
\end{eqnarray}
The total field $\hat{v}(\tau,x)$ can be expressed as,
\begin{equation}\label{eq:mComp}
    \hat{v}(\tau,x) =  \hat{v}_{0}(\tau,x) + \sum\limits_{i=1}^N \left(\hat{v}_i(\tau,x) + \hat{v}^*_{i}(\tau,x)\right) ,
\end{equation}
where the summation is over $N$ complex conjugate pairs of excitations. In general $N$ can be infinite but we can have a situation that only a finite number is relevant. For instance, it is possible that some of these excitations got trivial initial conditions. The  corresponding power-spectrum for $\hat{v}(\tau,x)$ in the momentum space can be written as
\begin{eqnarray} \nonumber
    P_{v}(k) &=& P_{v_0}(k) + \sum\limits_{i=1}^N P_{v_i}(k)+\sum\limits_{i=1}^N P_{v_i^*}(k)\\
    					&=& v_{0,k}(\tau) v_{0,k}^*(\tau) + \sum\limits_{i=1}^N  v_{i,k}(\tau) v_{i,k}^*(\tau) + \sum\limits_{i=1}^N  u_{i,k}(\tau) u_{i,k}^*(\tau),
\end{eqnarray}
where, we have used commutation relations \eqref{eq:comm1}.

Recalling that $z \sim a \chi\pr/\Hc$ and using relations $\hat{v}_{0}(\tau,x) = z \zeta_{0}(\tau,x)$, $\hat{v}_{i}(\tau,x) = z \hat{\zeta}_{i}(\tau,x)$, and $\hat{v}_{i,k}^*(\tau,x) = z \hat{\zeta}_{i}^*(\tau,x)$ we can write the power-spectra of the Fourier modes $\zeta_{0,k}$, $\zeta_{i,k}$ and $\zeta_{i,k}^*$ as
\begin{eqnarray}\label{m0Ps}
    \Delta_{\zeta_0}(k) = \frac{k^3}{2 \pi^2}\frac{1}{z^2} P_{v,0}(k) =  \frac{k^3}{2 \pi^2}\frac{1}{2 M_P^2 a^2\epsilon} (-\tau)\frac{\pi}{4} \vert H^{(1)}_{\nu_0} (-k \tau) \vert^2, \\
    \Delta_{\zeta_i}(k) = \frac{k^3}{2 \pi^2}\frac{1}{z^2} P_{v_i}(k) = \frac{k^3}{2 \pi^2}\frac{1}{2 M_P^2 a^2 \epsilon} (-\tau)\frac{\pi}{4} \vert H^{(1)}_{\nu_i} (-k \tau)\vert^2 e^{i(\nu-\nu^*)}, \\
     \Delta_{\zeta_i^*}(k) = \frac{k^3}{2 \pi^2}\frac{1}{z^2} P_{v_i^*}(k) = \frac{k^3}{2 \pi^2}\frac{1}{2 M_P^2 a^2 \epsilon} (-\tau)\frac{\pi}{4} \vert H^{(1)}_{\nu_i^*} (-k \tau)\vert^2 e^{i(\nu^*-\nu)}.
\end{eqnarray}
The total scalar curvature power-spectrum becomes
\begin{equation}
    \Delta_S = \Delta_{\zeta_0}(k) + \sum\limits_{i=1}^N\Delta_{\zeta_i}(k) + \sum\limits_{i=1}^N\Delta_{\zeta_i^*}(k).
\end{equation}

In order to test the power-spectrum with observations we need to evaluate the power-spectrum at the end of the inflation. To compute the behavior of the curvature power-spectrum, we can use the following behavior of Hankel functions at $k \tau \rightarrow 0$ limit (note that $\tau\to0_-$ limit corresponds to $t\to+\infty$ limit for the cosmic time $t$),
\begin{equation}\label{Hankel0}
	\lim_{k \tau \rightarrow 0} H^{(1)}_{\nu}(-k \tau) \approx \sqrt{\frac{2}{\pi}} e^{i \frac{\pi}{2}\left(\nu+\frac{1}{2}\right)} 2^{\nu - \frac{3}{2}} \frac{\Gamma(\nu)}{\Gamma\left({\frac{3}{2}}\right)} (-k \tau)^{\nu}\sim(-k\tau)^\nu.    
\end{equation}
The power-spectrum of massless excitation   \eqref{m0Ps} does not evolve after the horizon crossing which can be seen from the super-horizon behavior of the massless excitation using   \eqref{Hankel0}.
But for the massive case the situation is different. To ensure the massive excitations do not grow after the horizon crossing one should put a constraint  \eqref{mass-parabola} on the complex masses. In terms of Hankel function \eqref{Hankel0} in the  super-horizon limit where it behaves as $(-k \tau)^{3/2 - \nu_i}$ this constraint implies that $Re(\nu_i) \leq 3/2$.

\begin{figure}
	\centering
	\includegraphics[width=0.8\textwidth]{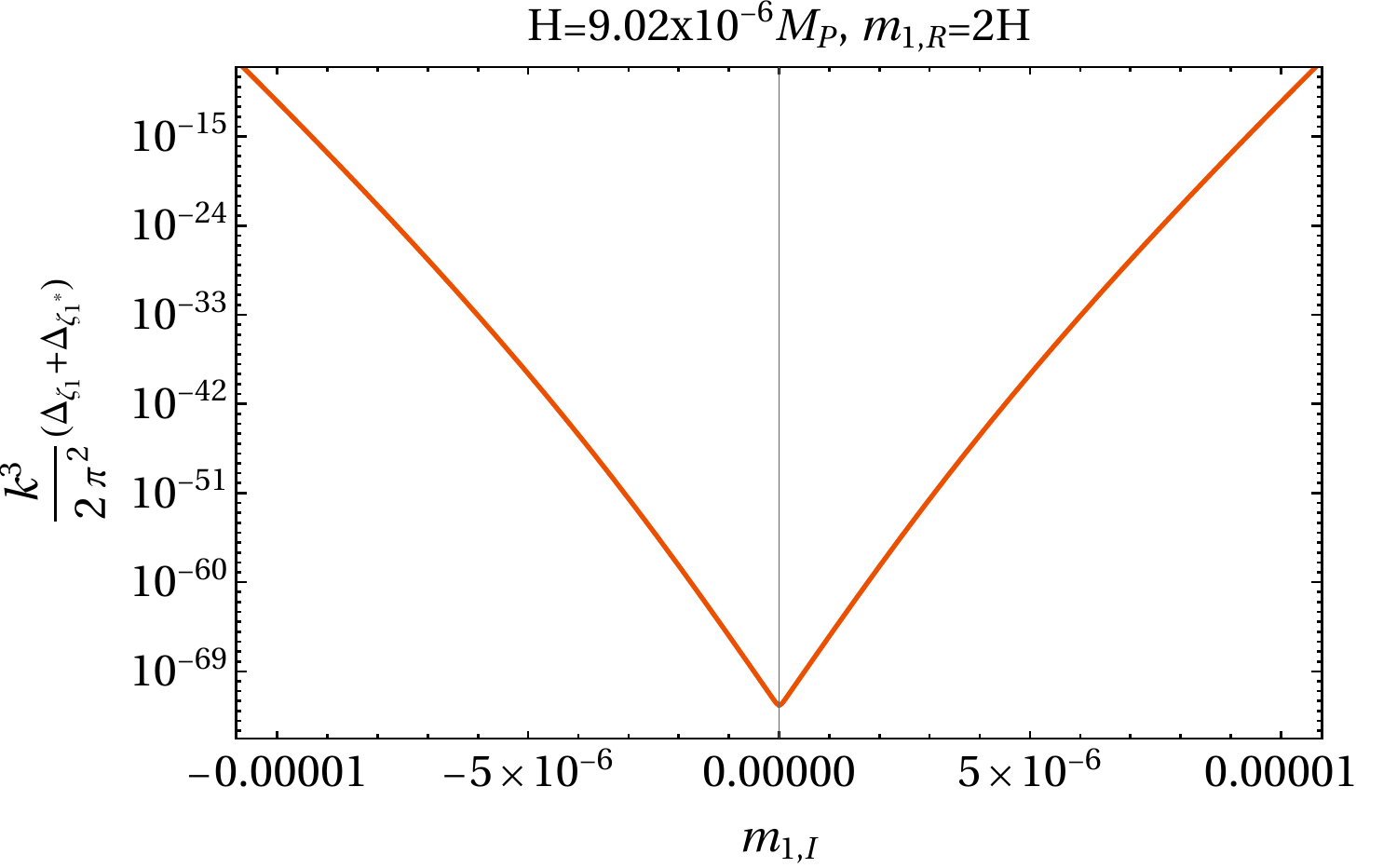}    
	\caption{\it  The dependence of the scalar power-spectrum due to massive excitations $(\Delta_{\zeta_1}(k)+\Delta_{\zeta_1^*}(k))$ in the case of one massive complex conjugate pair as a function of the imaginary part of mass $m_{1,I}$ (in units of $H$). Here $H = 9.02\times 10^{-6} M_P$, $m_{1,R} =2H$.}
	\label{fig:PsmR2H}
\end{figure}

Under our assumption that $\Mc \gg H$ and a natural expectation that the form-factor is a monotonic function (the latter does not have to be the case but simplifies greatly the proof of unitarity \cite{Koshelev:2021orf}) we can infer that the real part of the complex masses should be larger than $H$ such that $ m_{i,R}^2 > H^2$. Indexes $R,I$ will be used to designate the real and imaginary parts respectively. We can consider as an example one pair of complex conjugate excitations corresponding to $i=1$ and having masses $m_1$ and $m_1^*$. In Fig.~\ref{fig:PsmR2H} we plot the dependence of $(\Delta_{\zeta_1}(k)+\Delta_{\zeta_1^*}(k))$ as a function of the imaginary part of the mass $m_{1,I}$ for $H = 9.02 \times 10^{-6} M_p$ and $m_{1,R} = 2H$. We can see that for $m_1$ inside the mass parabola $\Delta_{\zeta_1}(k)$ and $\Delta_{\zeta_1^*}(k)$ are sufficiently small and the scalar power-spectrum $\Delta_S \sim \Delta_{\zeta_0}$. It is clear that the Planck constraint on scalar power-spectrum, $\Delta_S = 2.1 \times 10^{-9}$ \cite{Planck:2018jri} is satisfied. To find out the scale dependence of the scalar power-spectrum we can write
\begin{eqnarray}
    n_s - 1 = \frac{d \ln \Delta_S}{d \ln k} = \frac{d \ln \Delta_{\zeta_0}(k)}{d \ln k} + \frac{d}{d \ln k}  \ln \left(1+ \frac{\Delta_{\zeta_1}(k)}{\Delta_{\zeta_0}(k)} + \frac{\Delta_{\zeta_1^*}(k)}{\Delta_{\zeta_0}(k)} \right).
\end{eqnarray}
When $m_1$ is inside the mass parabola $\frac{\Delta_{\zeta_1}(k)}{\Delta_{\zeta_0}(k)} \sim \frac{\Delta_{\zeta_1^*}(k)}{\Delta_{\zeta_0}(k)} \ll 1$ we have  $n_s -1 \simeq \frac{d \ln \Delta_{\zeta_0}(k)}{d \ln k} = 2\eta - 6\epsilon$ which also satisfies the Planck bound on scalar power-spectrum scale dependence.

In general if masses are inside the mass parabola and amplitude of massive excitations are sufficiently suppressed they do not affect the scalar power-spectrum. Thus the amplitude of the scalar power-spectrum will be solely determined by the real massless excitation only, preserving inflationary predictions of a local single scalar field model. 
In the next Sections we are going to explore the effect of the new massive excitations on the scalar bi-spectrum. Even though the power-spectrum is not sensitive to masses inside the mass parabola, the bi-spectrum will be shown to be sensitive due to effects of integration on the way of computing non-Gaussian correlation functions.

\section{Three-point correlation functions}\label{sec:3pt}
To compute the three-point correlation functions or the bi-spectrum we use the In-In formalism where the three-point correlation functions can be evaluated as \cite{Maldacena:2002vr,Chen:2006nt},
\begin{eqnarray}\label{eq:3pt-master}
    \left\langle \zeta(t,x) \zeta(t,y) \zeta(t,z)\right\rangle = \left\langle\left(\int_{t_0}^t i~ dt^{\prime} H_{int}(t^{\prime})\right) \zeta(t,x) \zeta(t,y) \zeta(t,z)\right\rangle \\ \nonumber - \left\langle\zeta(t,x) \zeta(t,y) \zeta(t,z) \left(\int_{t_0}^t i~ dt^{\prime} H_{int}(t^{\prime})\right)\right\rangle .
\end{eqnarray}
Here $\zeta(t,x) = v(t,x)/z$ and here we recall that $v=v_0+\sum_i v_i$ is the non-local Mukhanov-Sasaki variable containing contributions of a real excitation for which we use the index $0$ and new complex mass excitations enumerated by $i$. Obviously $\zeta$ also becomes a linear superposition of $\zeta_0$ and $\zeta_i$-s. $H_{int}$ is the interaction Hamiltonian and $t_0$ is the time at the beginning of inflation.
       
The interaction Hamiltonian $H_{int}$ consists of terms that originate from the third variation of the action \eqref{eq:Lag-full}. Variation of the local and non-local part of the action gives rise to two respective sets of terms. Bi-spectrum due to the terms that originate from the local part of the action is well studied in the literature \cite{Maldacena:2002vr,Senatore:2009gt}. The third variation of the non-local part of the action  would give rise to three kinds of terms. The first kind of terms is those that are suppressed by higher order in slow-roll parameters. The second kind of terms is those that would have more than two spatial derivatives which will make them soft in the super-horizon limit \cite{Maldacena:2002vr,Senatore:2009gt}. We can safely ignore these two kinds of terms. {Then there is a third kind of terms. These are those  that are not slow-roll suppressed and do not have more than two spatial derivatives acting on them, so we should include these terms in the evaluation of the bi-spectrum.} These latter terms are still suppressed by factors of $1/\Mc^{2n}$ though. {If the inflaton mass is much smaller than $\Mc$} we expect that their effect may be negligible \cite{Koshelev:2022bvg}, but a detailed analysis of these operators is necessary. We wish to address this in future works. Anyway, we already have a major new ingredient in our model which is the new states with complex masses. In what follows we restrict ourselves by studying the effect of non-locality on the three-point correlation functions through the appearance of these new excitations. 

According to the above considerations we are going to evaluate the bi-spectrum originating from local single field third order interaction terms \cite{Maldacena:2002vr} only. The corresponding terms are as follows
\begin{eqnarray}\label{op-1}
	\mathcal{O}_1 &=& \frac{1}{M_P^2} a^3  \epsilon^2  \dot{\zeta}^2 \zeta,\\ \label{op-2}
	\mathcal{O}_2 &=& \frac{1}{M_P^2} a  \epsilon^2  \zeta \partial_k \zeta \partial^k \zeta, \\ \label{op-3}
	\mathcal{O}_3 &=& -\frac{2}{M_P^2} a^3 4 \epsilon^2  \dot{\zeta} \partial_k \zeta \partial^k (\partial^{-2} \dot{\zeta}).
\end{eqnarray}
The interaction Hamiltonian can be constructed out of these operators as follows
\begin{equation}
	H_{int} = - \int d^3 x \left(\mathcal{O}_1(t,x)+\mathcal{O}_2(t,x)+\mathcal{O}_3(t,x)\right).
\end{equation}
We will be interested in evaluating the bi-spectrum in the momentum space, and integrals in \eqref{eq:3pt-master} can be written  in terms of Wightman functions \cite{Maldacena:2002vr,Collins:2011mz,Chen:2006nt}. In a local single-field slow-roll inflation Wightman function for scalar perturbations can be defined as
\begin{equation}
    G_0^{>} (t,x; t^{\prime},y) = \langle 0 \vert \hat{\zeta}_0(t,x) \hat{\zeta}_0(t^{\prime},y)\vert 0 \rangle = \int \frac{d^3 k }{(2 \pi)^3} e^{i \vec{k}. (\vec{x}-\vec{y})} G^{>}_{0,k} (t,t^{\prime}),
\end{equation}
where as explained above $\hat{\zeta}_0(t,x) = \hat{v}_0(t,x)/z$. In the de Sitter limit where $\nu = 3/2$, the Wightman function in momentum space can be written as
\begin{equation}\label{eq:wigtmannM0}
	G^{>}_{0,k} (t,t^{\prime}) = \frac{H^2}{4 \pi M^2_{Pl} \epsilon} \frac{1}{k^3} (1+i k \tau) (1 - ik \tau^{\prime}) e^{-ik (\tau - \tau^{\prime})}.
\end{equation}
 We also define another Wightman function with reverse ordering of the fields which is just a complex conjugate of the previous one 
\begin{equation}\label{eq:WightmanM0cc}
	G^{<}_{0,k} (t,t^{\prime}) = \frac{H^2}{4 \pi M^2_{Pl} \epsilon} \frac{1}{k^3} (1-i k \tau) (1 + ik \tau^{\prime}) e^{ik (\tau - \tau^{\prime})}.
\end{equation}

 In the non-local scenario with curvature perturbation field $\hat{\zeta}(x,t) = \hat{v}(t,x)/z$ being a superposition of many individual curvature perturbations of different masses we can define the corresponding Wightman function as,
\begin{eqnarray}\label{eq:WightmanNL-1}
    G^{>} (t,x; t^{\prime},y) = \langle 0 \vert \hat{\zeta}(t,x) \hat{\zeta}(t^{\prime},y)\vert 0 \rangle = \int \frac{d^3 k }{(2 \pi)^3} e^{i \vec{k}.(\vec{x}-\vec{y})} G^{>}_{k} (t,t^{\prime}).
\end{eqnarray}
Dividing the non-local Mukhanov-Sasaki variable \eqref{eq:mComp} by $z$ and using commutation relations   \eqref{eq:comm1} we can write 
\begin{eqnarray}\label{eq:wightTotal}
    G^{>}_{k} (t,t^{\prime}) = G^{>}_{0,k}(t,t^{\prime}) + \sum\limits_{i=1}^N \left(G^{>}_{i,k}(t,t^{\prime})+\tilde{G}^{>}_{i,k}(t,t^{\prime})\right).
\end{eqnarray}
Here $N$ as explained above can in principle be infinite. But we can start by computing an effect of a finite number of contributions to the Wightman function. The $i$-th pair of Wightman functions due to the $i$-th pair of massive excitations with complex conjugate masses can be written as
\begin{eqnarray}\label{WightmanM}
	G^{>}_{i,k}(t_1,t_2) = \frac{H^2}{8 \epsilon M_p^2} \pi (-\tau_1 \tau_2)^{3/2}
	e^{i \frac{\pi}{2}\left(\nu_i - \nu_i^*\right)}H_{\nu_i}^{(1)}(-k\tau_1) H_{\nu^{*}_i}^{(2)}(-k\tau_2),  \\
	\tilde{G}^{>}_{i,k}(t_1,t_2) =\frac{H^2}{8 \epsilon M_p^2} \pi (-\tau_1 \tau_2)^{3/2} e^{i \frac{\pi}{2}\left(\nu_i^* - \nu_i\right)}H_{\nu_i^*}^{(1)}(-k\tau_1)H_{\nu_i}^{(2)}(-k\tau_2).
\end{eqnarray}
As mentioned above the Wightman functions with reverse ordering of the fields can be computed by complex conjugation. Additionally we need a time derivative of the Wightman functions with respect to the second argument to evaluate the bi-spectrum for operators \eqref{op-1} and \eqref{op-3}. We will denote time derivative of any Wightman function $G^{>}(t,t\pr)$ with respect to the second time argument $t\pr$ as $\dot{G}^>(t,t\pr)$.

To evaluate the bi-spectrum in the momentum space we need to compute time integrals involving the Wightman functions and its time derivatives. To illustrate the integrals with a non-local Mukhanov-Sasaki variable we consider the operator \eqref{op-1}. An integral corresponding to it can be written as
\begin{multline}\label{op-1Int} 
	B^{(1)}_{S}(k_1,k_2,k_3) \propto i \int dt^{\prime}  Im\left[\dot{G}_{0,k_1}^{>}(t,t^{\prime})\dot{G}_{0,k_2}^{>}(t,t^{\prime}){G}_{0,k_3}^{>}(t,t^{\prime}) \right. \\ 
	\left. +  \sum\limits_{i=1}^N \left\{\dot{G}_{0,k_1}^{>}(t,t^{\prime})\dot{G}_{0,k_2}^{>}(t,t^{\prime})\left({G}_{i,k_3}^{>}(t,t^{\prime})+\tilde{G}_{i,k_3}^{>}(t,t^{\prime})\right)\right. \right. \\ 
	\left. \left.
	+G_{0,k_1}^{>}(t,t^{\prime})\dot{G}_{0,k_2}^{>}(t,t^{\prime})\left(\dot{G}_{i,k_3}^{>}(t,t^{\prime})+\dot{\tilde{G}}_{i,k_3}^{>}(t,t^{\prime})\right) \right. \right. \\ 
	\left. \left.  + \dot{G}_{0,k_1}^{>}(t,t^{\prime})\left(\dot{G}_{i,k_2}^{>}(t,t^{\prime})+\dot{\tilde{G}}_{i,k_2}^{>}(t,t^{\prime})\right)\left({G}_{i,k_3}^{>}(t,t^{\prime})+\tilde{G}_{i,k_3}^{>}(t,t^{\prime})\right) \right. \right. \\ 
	\left. \left. +
	 {G}_{0,k_1}^{>}(t,t^{\prime})\left(\dot{G}_{i,k_2}^{>}(t,t^{\prime})+\dot{\tilde{G}}_{i,k_2}^{>}(t,t^{\prime})\right)\left(\dot{G}_{i,k_3}^{>}(t,t^{\prime})+ \dot{\tilde{G}}_{i,k_3}^{>}(t,t^{\prime})\right)\right. \right. \\ 
	\left. \left. + \left(\dot{G}_{i,k_1}^{>}(t,t^{\prime})+\dot{\tilde{G}}_{i,k_1}^{>}(t,t^{\prime})\right)\left(\dot{G}_{i,k_2}^{>}(t,t^{\prime})+\dot{\tilde{G}}_{i,k_2}^{>}(t,t^{\prime})\right) \right. \right. \\ 
	\left. \left.
    \times \left(\dot{G}_{i,k_3}^{>}(t,t^{\prime})+ \dot{\tilde{G}}_{i,k_3}^{>}(t,t^{\prime})\right)
    \right\} + \text{cyclic permutations } (k_1\to k_2\to k_3)
	\right].
\end{multline}
Here we have used the $\propto$ sign as we have not explicitly written factors containing slow-roll parameters or $H$, or $M_P$, as well as the momentum conservation delta-function $\delta^{(3)}(\vec{k}_1+\vec{k}_2+\vec{k}_3)$. Contributions of other operators in  \eqref{op-2} and \eqref{op-3} can be written using corresponding combinations of Wightman functions and their derivatives. Using this technique we will compute the three-point correlation functions and compare them to most up to date observations from Planck \cite{Planck:2018jri}. 

At this stage we see that actual computations use masses of excitations as the main input.
As explained in the Introduction it is a very difficult task to find masses in question even if a form-factor is given explicitly. We instead can treat mass values as parameters and analyze observational consequences based on its value. Such an approach can obviously narrow a class of possible form-factors by excluding those resulting in excitations with masses producing effects not supported by current observations.

Before continuing to the computation of the bi-spectrum we emphasize again that even though there are infinitely many excitations corresponding to operator $\mathcal{P}(\Box)$ in \eqref{eq:nlOp} they only produce cross-terms within a pair of excitations sharing complex conjugate masses.
Different pairs do not produce cross-terms upon computations due to vanishing commutators of operators as follows from (\ref{eq:comm1}).
It then sounds reasonable as a first step  to reveal effects of non-locality on the three-point correlation functions by  considering only one pair of excitations with complex masses. This effectively means that $N=1$ in summations over different contributions.

From limit \eqref{Hankel0} we can see that for masses inside the mass parabola \eqref{mass-parabola}, the closer the mass of the excitation to the real axis is, the more suppressed contribution to the power-spectrum it gives. There should be only one relevant real mass excitation having mass $m_0$ which matches to the standard inflaton and is nearly massless. It cannot be excluded from any consideration as otherwise known predictions will not be reproduced without a heavy fine-tuning. This excitation moreover is expected to be the lightest (in the absolute value). But there should not be other real mass excitations as they will become physical ghosts.

Thus we proceed by taking into account one real mass excitation with the mass $m_0$ and one pair of complex conjugate mass excitations with masses $m$ and $m^*$ which as argued before should be heavier (in absolute value) than $m_0$. The question to study is the influence of such extra excitation on the bi-spectrum as a function of a position of this mass on the complex plane with respect to the border of the parabolic stability domain which is intrinsically determined by the Hubble parameter during inflation. From limit (\ref{Hankel0}) masses obviously cannot be outside of this region as even the power-spectrum predictions will start easily contradicting observations. This reduces the question to how close from inside to the border of the parabolic domain of classically stable excitations masses still can be present. Moreover, due to significant difficulties in using analytic computations we resort to numeric methods which are discussed in details in the next Section.


\section{Numerical computation of three-point correlation functions}\label{sec:numerical}

\subsection{Computation scheme}

The main difficulty to evaluate the integrals of the form \eqref{op-1Int} which involve Hankel functions with complex indices is the fact that there are no known compact analytic representations of Hankel functions on the entire complex plane for an arbitrary complex index and argument. There are many developments in this direction and an interested reader can consult \cite{Olver1,Olver2,Paknys,Sastry,Dunster1}. For integrals involving Hankel functions one can refer to \cite{Bailey1}. But in all these instances either a purely imaginary index was considered or the analysis was done in a restricted region on the complex plane of the argument. In our case these limitations have to be bypassed and in order to get results we compute the bi-spectrum numerically.

While performing numerical integration of the form of \eqref{op-1Int} one faces the challenge to make the integrals convergent when $\tau \rightarrow -\infty$ as they are heavily oscillating. In the original work \cite{Maldacena:2002vr}, the time variable $\tau$ was continued to the imaginary plane such that the oscillating behavior gets exponentially suppressed. In \cite{Chen:2009zp}, the computation of the three-point correlation functions was done numerically in the context of quasi-single-field inflation. In that paper to achieve convergence a Wick rotation was used analytically continuing $\tau  \rightarrow i \tau$.\footnote{For other recent developments to deal with the numerical computations of non-Gaussianities one can refer to \cite{Pinol:2023oux,Dias:2016rjq}} Recently methods based on Cesaro summation for oscillatory integrals are used in the context of non-Gaussianities \cite{Junaid:2015hga} and this method can be expressed as a nested integral 
\begin{equation}\label{eq:cesaro}
	I_{\text{Cesaro}} = \int_{-\infty}^{\tau_0}  d \tau^{\prime} f(\tau^{\prime}) = \lim_{\tau \rightarrow -\infty} \frac{1}{\tau_0 - \tau} \int_{\tau_0}^{\tau} d\tau^{\prime} \int_{\tau_0}^{\tau^{\prime}} d \tau^{\prime \prime} I(\tau^{\prime \prime}).
\end{equation}
To avoid the complexity of the nested integrals one can use an equivalent method of Riesz summation where an integral can be expressed as \cite{Tran:2022euk}
\begin{equation}\label{eq:riesz}
	I_{\text{Riesz}}	= \int_{-\infty}^{\tau_0} d \tau^{\prime} f(\tau^{\prime}) = \lim_{\tau \rightarrow -\infty} \int_{\tau}^{\tau_0} d \tau^{\prime} \left(1 - \frac{\tau^{\prime} - \tau_0}{\tau - \tau_0}\right)^{n} f(\tau^{\prime}),
\end{equation}
where $n$ is a non-negative integer. If the non-oscillatory part of the integrand is behaving as $\tau^m$ for some positive $m$ then one should choose $n > m$. These two schemes of Cesaro and Riesz summation are equivalent to each other. They are also equivalent to the Wick rotation method and to the method of continuation of $\tau$ to the complex plane, but the convergence speed is better for Cesaro and Riesz schemes  \cite{Tran:2022euk}. Riesz summation has one single integration and it is easier to evaluate if the analytic form of the Wightman functions is known. Since in our scenario the analytic form for the Wightman functions is known in terms of Hankel functions, we use the Riesz summation method to evaluate the integrals.

\subsection{Numerical results}

We are interested to analyze the effect of the complex masses on the bi-spectrum at the CMB scales. Considering the current bounds from Planck 2018 \cite{Planck:2018jri} on pivot scale $k_{\star}$, we fix the relevant parameters as follows
\begin{equation}
		H = 9.02 \times 10^{-6} M_P; ~ \epsilon = 10^{-3};~ \eta = 0.03;~ M_P = 1;~ k_{\star} = 0.05 h Mpc^{-1}.
\end{equation}
To analyze the effect of complex masses on the three-point correlation functions we select two benchmark points for the real part of the complex mass, $m_R = 2 H$ and $m_R= 10 H$. {Here as we consider only one pair of excitations we drop the index $i$ from the mass notation $m_i$. Also we recall that we denote the real and imaginary parts of the mass as $m_R$ and $m_I$ respectively. The index of Hankel function will be denoted with $\nu$.} With these two choices of $m_R$ the maximum allowed value for the complex part of the mass $m_I$ respecting mass parabola condition   \eqref{mass-parabola} are $m_I^{max} = 1.2 H$ and $m_I^{max} = 1.48 H$ respectively. We will see shortly that maximal values of $m_I$ would not be reached though.  

In order to use the Riesz summation scheme \eqref{eq:riesz} the choice of parameter $n$ is an important step. As mentioned earlier if the non-oscillatory part behaves as $\tau^m$ then one needs to choose $n > m$. We have checked that in our case the numerical results of the integration do not change for $n > 20$ and to be on the safe side we set $n = 40$.

To evaluate the amplitude of the bi-spectrum we use the following definition of $f_{NL}$ \cite{Planck:2019kim, Byrnes:2010ft},
\begin{equation}
    f_{NL} = \frac{5}{6} \frac{B_{S}(k_1,k_2,k_3)}{P_{S}(k_1)P_{S}(k_2)+P_{S}(k_1)P_{S}(k_3)+P_{S}(k_2)P_{S}(k_3)},
\end{equation}
where $B_{S}(k_1,k_2,k_3)$ is the three-point correlation function and $P_{S}(k_i) = \frac{2\pi^2}{k_i^3} \Delta_{S}$ with $\Delta_{S}$ being normalized to the Planck power-spectrum normalization, i.e. $\Delta_{S} = 2.1 \times 10^{-9}$. Our goal is to see how the obtained values for non-Gaussianity parameter constrain possible values of the complex masses of excitations while satisfying the constraint on $f_{NL}$ from Planck \cite{Planck:2019kim}. The latter tells that $\vert f_{NL}^{eq,sq}\vert < 10$, where ``eq" and ``sq" labels refer to equilateral and squeezed limit configuration of non-Gaussianity respectively. {The equilateral limit corresponds to $k_1 = k_2 = k_3$ and the squeezed limit corresponds to $k_1 \ll k_2 \sim k_3$.}

In general the integrals of interest contain a product of  three Hankel functions with different indices $\nu_0$ and $\nu$, $\nu^*$ corresponding to massless and massive excitations respectively. Namely we encounter integrals as follows
\begin{eqnarray}\label{eq:intType}
    \mathcal{I} \propto \int d\tau (-\tau)^n H_{\nu_i}^{(1,2)}(-k_1 \tau) H_{\nu_j}^{(1,2)}(-k_2 \tau) H_{\nu_k}^{(1,2)}(-k_3 \tau),
\end{eqnarray}
 where $\nu_i$-s can be $\nu_0$, $\nu$, $\nu^*$, $\nu_0 \pm 1$, $\nu \pm 1$ or $\nu^* \pm 1$ depending on the interaction terms  \eqref{op-1}, \eqref{op-2}, or \eqref{op-3} to be computed.  With the benchmark points used in computations the largest contributing integral is the combination of $H_{\nu_0}^{(1,2)} H_{\nu_0}^{(1,2)} H_{\nu,\nu^*}^{(1,2)}$ along with it's derivative combinations. Integrals containing $H_{\nu_0}^{(2)} H_{\nu}^{(2)} H_{\nu}^{(2)}$ exactly cancel the contribution of integrals containing $H_{\nu_0}^{(2)} H_{\nu^*}^{(2)} H_{\nu^*}^{(2)}$, even though they can be individually large. This is also true for the integrals that contain the same combinations of Hankel functions but with time derivatives acting on them according to the operators \eqref{op-1} and \eqref{op-3}. All other integrals are small when the mass is inside the mass parabola.

The main results of our numerical computations can be summarized as follows:
\begin{itemize}
    \item For $m_R = 2H$, $m_I = 0.5 m_{I}^{max}$ the numerical computations give $f_{NL}^{sq} = -17.90$ which violates the Planck constraint on this parameter. This enhancement in $f_{NL}^{sq}$ can be attributed to the presence of massive excitations as it is known that $f_{NL}^{sq}$ due to the massless mode is small and satisfies Maldacena consistency relation.  Contribution of these massive excitations to  $f_{NL}^{eq}$  is small and  can be estimated to be the same as in the local single-field case. Thus for $m_R = 2H$, we get a constraint $m_I < 0.5 m_{I}^{max}$ in order to be consistent with Planck observations.
    \item For $m_R=2H, ~m_I = 0.48 m_{I}^{max}$ the amplitude is computed to be $f_{NL}^{sq} = -0.97$ which is within the Planck bounds. 
    \item For $m_R = 10H, ~ m_I = 0.52 m_{I}^{max}$ the numerical computation gives $f_{NL} = 8.41$ which is within Planck bounds. We have found that  for $m_R = 10H, ~ m_I > 0.53 m_{I}^{max}$ the Planck bounds get violated.
    \item For the choice of $m_R = 10H,~ m_I = 0.5 m_{I}^{max}$, we obtain $f_{NL}^{sq} = 0.45$ which is well within the Planck bounds.
    \item With the choices of $m_R$ and $m_I$ discussed above, $f_{NL}^{eq}$ remains the same as in the local single-field inflation. So the equilateral limit of bi-spectrum is not sensitive to the presence of the new complex mass excitations.
    \item Finally, for $m_I>m_I^{max}$ we obtain that $f_{NL}^{sq}$ grows much steeper quickly gaining extra orders of magnitude and thus confirming that the stability region is indeed a special domain for possible values of masses of new states.
\end{itemize}

\section{Discussion and outlook}

In this paper we have explored cosmic inflation in a non-local scalar field setup where the scalar field serves as the inflaton field and the gravity is described by standard Einstein-Hilbert term. In our approach the non-locality is introduced as a form-factor $\mathcal{F}(\Box)$ which being a function of a d'Alembertian operator manifests the presence of infinite derivatives. Starting from the non-local action   \eqref{eq:Lag-full} we first canonically normalize the system leading to a potential with non-local argument. We then systematically perturb the system and using slow-roll conditions we arrive at a non-local generalization of the Mukhanov-Sasaki equation. The non-locality in Mukhanov-Sasaki equation results in the presence of an infinite number of degrees of freedom. By using the Weierstrass product decomposition method we demonstrate that this non-local system generates a real nearly massless excitation similar to the local Mukhanov-Sasaki variable along with new excitations with complex masses which come in complex conjugate pairs. The non-local Mukhanov-Sasaki variable becomes a sum of the massless excitation and pairs of excitations with complex masses. Upon computing the power-spectrum of the curvature perturbations with the non-local Mukhanov-Sasaki variable we observe the direct correspondence with the constraint on the complex masses coming from   condition \eqref{mass-parabola} which describes a parabolic region of allowed complex masses values \cite{Koshelev:2020fok}.  Excitations with masses inside this region classically decay but those outside do grow and subsequently ruin the inflationary background and the power-spectrum constraints from Planck \cite{Planck:2018jri}.

As one of the main result in the present paper we compute the amplitude of bi-spectrum $f_{NL}$ in equilateral and squeezed limit configurations in the presence of the new excitations with complex masses in the non-local Mukhanov-Sasaki variable. While these new excitations come in pairs with complex conjugate masses, effects of different pairs do not mix to commutation relations of respective quantum operators \eqref{eq:comm1}.
We proceed with taking into account a single pair of new states and perform the corresponding computations numerically using the Riesz summation method \cite{Peterson:2010mv}.
Our computations of $f_{NL}$ reveal additional stronger than simply a border of the mass parabola region constraints on the masses of new states.

Namely, for a single pair of massive excitations with the real part of the mass $m_R = 2H$, the imaginary part has to satisfy $m_I < 0.5 m_I^{max}$ in order to be consistent with Planck's bound on $f_{NL}^{sq}$. For another choice of the real part of the mass $m_R = 10H$ the Planck bounds suggest that the imaginary part of the mass has to satisfy $m_I < 0.53 m_I^{max}$. So in order to be consistent with Planck's bound on the non-Gaussianity the complex masses can not reside at the edge of the mass parabola region described by the condition \eqref{mass-parabola}, rather it needs to be about half deep inside of it. 
This outcome follows from the properties of integrals involved in computing the bi-spectrum. It is an interesting question for upcoming works to understand better physics behind this stronger restrictions.

We have also found that the equilateral limit of the bi-spectrum is not sensitive to the presence of the massive excitations at least for the above discussed values of $m_R$ and $m_I$. These observation form an apparent hint for observational bounds as the discussed limits will hopefully be measured more accurately in near future.

An important aspect to note is that there is no a well established transformation between Jordan and Einstein frames in non-local gravitational theories. The key issue is the presence of essential nonlinearities with respect to the metric in the action. The bi-spectrum in a non-local $R^2$ inflation was studied before in \cite{Koshelev:2020foq,Koshelev:2022bvg} but it is difficult to make a direct comparison of our results. However we can make a correspondence between those models based on obtained observational signatures. In particular we see that the Jordan frame model enhances both the equilateral and the squeezed limits (in fact also the orthogonal limit as well) of the  bi-spectrum compared to the single-field slow-roll scenario because of the occurrence of new operators in the Lagrangian due to the non-locality. However this enhancement happens only if the mass of the inflaton is very close to the scale of non-locality which is not the case under our current assumptions.

In the present paper starting from a non-local scalar field theory in an Einstein frame and assuming that the inflaton mass is much less than the scale of non-locality, we find that only the squeezed limit of the bi-spectrum is enhanced leaving the equilateral limit the same as in a single-field slow-roll inflation. In our case the reason for the enhancement is the non-local nature of the Mukhanov-Sasaki variable which introduces new excitations with complex masses and hence the curvature perturbation gets contribution from a massless excitation and new massive excitations with complex masses. In the present paper we omitted operators that can originate from the non-local part of the Lagrangian thanks to our assumption that the inflaton mass is much smaller than the non-locality scale and this is why our results look different compared to  \cite{Koshelev:2020foq,Koshelev:2022bvg}. 

Computations of scalar perturbations for non-local scalar field inflation setup \eqref{eq:Lag-full} reveal the non-local nature of the curvature perturbations by the introduction of new complex mass excitations. Constraints on the scalar power-spectrum suggest that the complex masses should reside inside a parabola shaped region \eqref{mass-parabola}. Assuming the inflaton mass to be much smaller than the non-locality scale, computation of the bi-spectrum with operators originating from the local part of the Lagrangian puts further constraints on these masses in order to be consistent with the observations. Effectively the parabola shaped region is narrowed. However, a further analysis of the operators originating from the non-local part of the Lagrangian is necessary to analyze the bi-spectrum in a wider range of parameter space. This analysis is also required for a better understanding of the Einstein-Jordan frame correspondence in non-local theories. We aim to address these questions in near future.

Other important direction for further study in the framework of the present paper is an implementation of the bootstrap program for cosmological correlation functions when new BIS-s with complex masses are present in the perturbation spectrum.

\acknowledgments

We thank Anish Ghoshal for his initial collaboration in this project. We would like to thank Anna Tokareva for numerous fruitful discussions. We also thank Dong-Gang Wang and Masahide Yamaguchi for discussions on various related questions.


\bibliographystyle{JHEP}
 \bibliography{biblio}






\end{document}